\documentclass[aps,preprint]{revtex4}
\usepackage{epsfig}
\usepackage{amsmath}
\usepackage{epsfig}
\begin{document}

\title
{\bf New features of scattering from a one-dimensional non-Hermitian (complex) potential}    
\author{Zafar Ahmed}  
\email{zahmed@barc.gov.in}
\affiliation{Nuclear Physics Division, Bhabha Atomic Research Centre,
Mumbai 400 085, India}

\date{\today}


\begin{abstract}
For complex one-dimensional potentials, we propose the asymmetry of both reflectivity and transmitivity
under time-reversal: $R(-k)\ne R(k)$ and $T(-k) \ne T(k)$, unless the potentials are real or PT-symmetric. 
For complex PT-symmetric scattering potentials, we propose that $R_{left}(-k)=R_{right}(k)$ and $T(-k)=T(k)$.
So far, the spectral singularities (SS) of a one-dimensional non-Hermitian scattering potential are
witnessed/conjectured to be at most one. We present  a new non-Hermitian parametrization of Scarf II potential
to reveal its four new features. Firstly, it displays the just acclaimed (in)variances.  Secondly, it can support 
two spectral singularities at two pre-assigned real energies ($E_*=\alpha^2,\beta^2$) either in $T(k)$  or in $T(-k)$, 
when $\alpha\beta>0$. Thirdly, when $\alpha \beta <0$ it possesses one SS in $T(k)$ and the other in $T(-k)$.
Fourthly, when the potential becomes PT-symmetric $[(\alpha+\beta)=0]$, we get $T(k)=T(-k)$, it possesses a unique 
SS at $E=\alpha^2$ in both $T(-k)$ and $T(k)$. Lastly, for completeness, when $\alpha=i\gamma$ and $\beta=i\delta$,
there are no SS, instead we get two negative energies  $-\gamma^2$ and $-\delta^2$ of the complex PT-symmetric 
Scarf II belonging to the two well-known branches of discrete bound state eigenvalues and no spectral singularity 
exists in this case. We find them as $E^{+}_{M}=-(\gamma-M)^2$ and $E^{-}_{N}=-(\delta-N)^2$;
$M(N)=0,1,2,...$ with $0 \le M (N)< \gamma (\delta)$.
\end{abstract}

\maketitle

A particle subject to a potential in the Schr{\"o}dinger equation is what lies beneath the foundation of 
non-relativistic quantum mechanics. A variety of potentials in various situations keep throwing 
new results  where proofs are elusive (or incomplete) and experimental verification could be 
challenging. In the non-Hermitian domain, for over a decade the complex PT-symmetric potentials [1] 
have attracted a large number of investigations of both types theoretical [2] and experimental [3]. A Hamiltonian
which is invariant under the joint transformation of Parity ($P: x \rightarrow -x$)  and time-reversal 
($T:i \rightarrow -i$) is called PT-symmetric. Later these new Hamiltonians have been covered under the 
more general concept of pseudo-Hermiticity [4]. However, the language of PT-symmetry is physically more 
appealing.

The present work brings out new features of scattering from a one-dimensional complex potential.

Earlier, for a general non-Hermitian (complex) scattering potential it has been proved that [5]
\begin{equation}
R_{left} \ne R_{right} \quad {\mbox and} \quad T_{left}= T_{right}.
\end{equation}
The reflectivity turns out to be symmetric: $R_{left}=R_{right}$, if the complex potential 
is spatially symmetric. For complex PT-symmetric potentials which are essentially spatially asymmetric it is found that if the 
particle enters from the side where potential is absorptive ($\Im(V(x))<0$), then the reflectivity
($R(E)$) is normal $<1$. But if it enters from the other side, $R(E)$ would be anomalous ($>1$) [7] for some
domain of energy. This phenomenon is called left/right-handedness of the reflectivity for a complex PT-symmetric
scattering potential.

In this Letter, we claim that the proof for (1) (see Eqs.(7,9) in [5]) subsequently
for a complex non-Hermitian potential also yields the asymmetry of both reflectivity and  transmitivity
under time-reversal as
\begin{equation}
R(-k) \ne R(k), \quad T(-k) \ne T(k).
\end{equation}
The indicated proof of (2) is not sufficient to rule out the exclusive PT-symmetric scattering potentials
for which we conjecture
\begin{equation}
R_{left}(-k)= R_{right}(k), \quad T(-k) = T(k).
\end{equation}
We claim that models of complex PT-symmetric scattering potentials discussed so far [6-9] and others  indeed conform to these 
new invariances (3), yet a proof is welcome. A similar lacunae has been faced earlier wherein complex periodic PT-symmetric potentials
could admit [10] real energy-bands yet a proof is desired. The unique signature/feature of complex PT-symmetric potentials even 
in scattering is a desideratum. In this regard, the conjecture (3) along with the left/right-handedness (1) of the reflectivity 
could be the unique feature/signature of scattering from a complex PT-symmetric potential.

One feature of scattering from a complex potential which is now  important [8] though present (see Fig~2. in [7]) got overlooked 
then. This feature is of an anomalous $(>>1)$ peak in both transmission ($T(E)$ and in (left) 
reflection co-efficients. Recently, the positive real discrete energy ($E=E_*>0$) at which both $R(E)$ and $T(E)$ become infinity has been well 
investigated as spectral singularity or zero width resonance [8]. These are real discrete energies where the Jost functions 
become linearly dependent. It is well to recall that in Hermitian quantum mechanics the physical 
energy-poles of $T(E)$ and $R(E)$ if real, represent the bound states ($E_n < 0$) and resonances (meta-stable states) if complex: 
${\cal E}_n-i\Gamma_n/2$ with ${\cal E}_n>0$. 

The spectral singularities of a non-Hermitian one-dimensional potential have also been discussed 
earlier in connection with the super-symmetric quantum mechanics [9]. However, seeing [8] them as positive energy discrete poles of $T(E)$
and $R(E)$ is more transparent and also it connects well to their experimental realization in wave propagation experiments [11]. The role of 
spectral singularity in the completeness of the bi-orthogonal basis has been well debated [12]. Spectral singularities seem to have 
been known in the theory of differential equations with complex and variable co-efficients (see Refs. in [8,9]). 

The solvable Scarf II can be expressed as
\begin{equation}
V(x)=V_1 \mbox{sech}^2 x+ V_2 \mbox{sech} x \tanh x
\end{equation}
this can be complexified by taking $V_2=iU$.
This complexification has contributed quite considerably in the complex PT-symmetric quantum mechanics. This is the first exactly solvable 
model of complex PT-symmetric potential for: scattering [7], discrete spectrum (real and complex-conjugate)[13], and the spectral singularity [14]. 
It has also helped in various other investigations [15]. An explicit condition involving $V_1$ and $V_2$ could be derived along with a 
simple explicit expression for the spectral singularity [14]. These  results have been found  by analytically continuing the well known [16,17] expressions 
of transmission and reflection amplitudes of the Hermitian Scarf II potential. Later, in an update on the complex PT-symmetric Scarf II 
potential (4)  a part (when $V_1<0$) of these results on the spectral singularity [14] have been re-derived [18].

So far, for one-dimensional non-Hermitian potentials the spectral singularity has been witnessed/conjectured to be at most one [8,9]. In this Letter,
we wish to reveal existence of two spectral singularities in (4). We present a new parametrization for $V_1$ and $V_2$ to find $E_{*1}$
and $E_{*2}$ in terms four real parameters $m,n$ (integer) and $\alpha, \beta$. The fixed potentials with a unique SS will also occur here 
as special cases.

For the Scarf II potential
\begin{equation}
V(x)=(B^2-A^2-A)\mbox{sech}^2x + B(2A+1) \mbox {sech}x \tanh x,
\end{equation}
Let $2\mu=1=\hbar^2$ and $k=\sqrt{E}$, where $E$ is the energy. Following [16,17] we can write the transmission amplitude for (5) as
\begin{eqnarray}
&t_{A,B}(k)=\frac{\Gamma[-A-ik] \Gamma[1+A-ik] \Gamma[1/2-iB-ik] \Gamma[1/2+iB-ik]}{\Gamma[-ik] \Gamma[1-ik] \Gamma^2[1/2-ik]},
\\ \nonumber
&r_{A,B}(k)=t_{A,B}(k)\left [\frac{\cos \pi A \sin \pi B}{\cosh \pi k}+ i \frac{\sin \pi A \cos \pi B}{\sinh \pi k} \right ].
\end{eqnarray}
The transmitivity $T(E)=|t(k)|^2$ and the reflectivity $R(E)=|r(k)|^2$.
We have rederived (6) to find that for (5)
\begin{eqnarray}
&&t_{left}(k)=t_{A,B}(k), r_{left}(k)=r_{A,B}(k)\\ \nonumber
&&\mbox{and} \quad t_{right}(k)=t_{A,-B}(k), r_{right}(k)=r_{A,-B}(k).
\end{eqnarray}
For the Hermitian case both $A$ and $B$ are real and the equation (6) displays various invariances: $R(-k)=R(k), T(-k)=T(k)$; 
and $R_{left}=R_{right}, T_{left}=T_{right}$ as usual.
For a general non-Hermitian potential we have $A=A_1+iA_2, B=B_1+iB_2$, we can check that the acclaimed properties (2) are followed.
When $A$ is real and $B$ is purely imaginary, we get a complex PT-symmetric potential in Eq. (5), one can indeed verify 
the acclaimed (3) PT-symmetry of the reflectivity and transmitivity from (6) by the help of (7). 

Now we parametrize $A$ and $B$ in (5) and (6) to bring out two spectral singularities. Let 
\begin{equation}
A=-(m+1)+i\alpha, \quad, \quad m \in I^++\{0\},
\end{equation}
then  the second Gamma function in the numerator of (3) is $\Gamma[-m+(\alpha -k)i]$ which becomes infinite ($\Gamma(-m)=\infty$)
when $k=\alpha$ giving us the first spectral singularity: $E_{*1}=\alpha^2$.
Next, when we assign 
\begin{equation}
B=\beta+i(n+1/2),\quad  n \in I^++\{0\},
\end{equation}
this time the third Gamma function in the numerator of (3) is $\Gamma[-n+(\beta-k)i]$ which becomes infinite ($\Gamma(-n)=\infty$)
when $k=\beta$ giving us the second singularity: $E_{*2}=\beta^2$.

The potential(s) possessing these two spectral singularities is Scarf II as in (4) which by the help of (5) gives
a new parametrization of $V_1$ and $V_2$ as
\begin{eqnarray}
&&V_1(\alpha,\beta)=[\alpha^2+\beta^2-(m+1)^2-(n+1/2)^2+(m+1)]\\ \nonumber
&&+i[(2m+1)\alpha+(2n+1)\beta],\\ \nonumber 
&&V_2(\alpha, \beta) =-[(2n+1)\alpha+(2m+1)\beta]\\ \nonumber
&&+i[2\alpha \beta -(m+1)(2n+1)+(n+1/2)], m,n \in I^++\{0\}.
\end{eqnarray}

In the following, we discuss SS in terms of transmitivity since the relevant positive energy poles of $R(k)$ and $T(k)$ are common (see Eq. (6)).
In Figs.~1, the transmitivity is represented by the dark curve and the time reversed transmitivity by the faint curve.   
The following cases arise here:\\ \\
{\bf (1) Two spectral singularities (general non-Hermiticity):} When $\alpha \beta >0$ there exists SS at $k_{*1}=\alpha$ and $k_{*2}=\beta$ or $k_{*1}=-\alpha$ and $k_{*2}=-\beta$. Alternatively, we can state  that both spectral singularities $E_{*1}=\alpha^2$ and $E_{*2}=\beta^2$ will occur either in the transmitivity (see two peaks 
in the dark curve in Fig.~1(a)) or in the time reversed transmitivity (see two peaks in the faint curve Fig.~1(b)). \\ \\
{\bf (2) Single spectral singularity (non-Hermiticity):} When $\alpha \beta <0$ ($\alpha+\beta \ne 0$) and then SS exists at $k_*=\alpha$ or at $k_*=-\beta$. Alternatively, one of the spectral singularities $E_{*1}=\alpha^2$ or  $E_{*2}=\beta^2$ will occur in the transmitivity and the other one in the time-reversed transmitivity. In Fig.~1(c) see one peak in the dark curve and one in the faint curve.
When $\alpha=i\gamma, (\gamma >0)$, there exists one bound state eigenvalue at $E=-\gamma^2$ (see (11) below) and one 
SS at $E_*=\beta^2$. The SS occurs in the transmitivity (time-reversed transmitivity) when $\beta >0 (<0).$
\\ \\
{\bf (3) Single spectral singularity( Complex PT-symmetry):} When $\alpha+\beta=0$ and $m=n$ meaning the Scarf II is PT-symmetric as $\Im(V_1)=0=\Re (V_2)$ (see (10)), 
there exists a unique spectral singularity $E_{*}=\alpha^2$. Alternatively, in the case of PT-symmetry the unique spectral singularity occurs in {\it both}:
the transmitivity and  the time reversed transmitivity. In Fig.~1(d),  both the dark and the faint curves merge in one displaying a single peak.
The cases when $\Re (V_1)$ is positive or negative arise here:
However, in both cases $|\Im(V_1)| > |\Re(V_2)|+1/4$ (see [14]). It may be noted that the sign of $V_1$ is considered opposite in [14]. 
\\ \\
{\bf (4) Bound states and no spectral singularity (Complex PT-Symmetry):} If we take $\alpha=i\gamma$ and $\beta=i\delta$ then for  non-negative integral values of $m,n$ in (8-10), we get complex PT-symmetric Scarf II potential. The negative energies $E=-\gamma^2, -\delta^2$ will be two real discrete eigenvalues of (5) belonging to
two branches of bound state eigenvalues. No spectral singularity can exist in this case.

We derive these two branches of eigenvalues from the physical poles of $t_{A,B}(k)$ such that $\Im (k) >0$.
Let use  $A=-(m+1)-\gamma$ (see (8)) in (6) to find  the second Gamma function yields the physical poles as  $k=i(m+\gamma-M)$, where $M$ is non negative integer: 
$M=0,1,2,...<m+\gamma$. From this  physical pole, we get one branch of negative discrete spectrum as
\begin{equation}
E^{+}_{M}=-(\gamma+m-M)^2, \quad M=0,1,2,...<m+\gamma.
\end{equation}
Similarly, the the third Gamma function in (6) also yields physical poles as $k=i(\delta+n-N)$, where $N$ is non-negative integer such that $N=0,1,2,...<n+\delta$.
We get the second branch of the negative discrete spectrum as
\begin{equation}
E^{-}_{N}=-(\delta+n-N)^2, \quad N=0,1,2,...<n+\delta.
\end{equation}
These eigenvalue formulae which are derived here from the physical ($\Im (k)>0$) $k$-poles of the transmission amplitude (6) can be verified as
newly expressed forms of the energy eigenvalue formulae derived earlier [13,15] for the complex PT-symmetric Scarf II potential.

It is demonstrated well that physical poles of $t(k)$ or $r(k)$ yield the spectral singularities and the bound states when $\Im (k) =0$ 
and when $\Im (k) > 0$, respectively [8]. It needs to be emphasized that $m,n$ themselves are non-negative integral parameters to be chosen along with 
$\gamma$ and $\delta$ for fixing the potential (5) using (8-10). The energies $-\gamma^2$ and $-\delta^2$ will essentially belong to the bound state eigenvalues
$E^{+}_M$ (11) and $E^{-}_N$ (12), respectively. So for instance when $m=n=0$, these will be the ground state eigenvalues of the two branches.
When two branches of real discrete spectra exist, the complex PT-symmetric Scarf II  is devoid of spectral singularity. Remarkably, 
the spectral singularity is not a necessary feature of scattering from a complex PT-symmetric potential.
It turns out that the spectral singularity is not the necessary feature of a complex PT-symmetric scattering potential. In these cases the spectral singularity
is more probable when the imaginary part of the potential is dufficiently stronger that its real part.

The Table I, displays the up-date on various (in)variances in the scattering from one-dimensional potentials. The results $\{1\}$, $\{2\}$ [5] are already known. The result $\{3\}$ is a new proposal (see Eq.(2)). All these three results can be proved readily using the method of the Ref. [5]. For complex PT-symmetric potentials, for the conjecture in $\{4\}$ (see Eq. (3)) a proof is welcome. It, however, along with the left/right-handedness of the reflectivity endows the complex PT-symmetric scattering potentials a unique signature.  

We believe that the present results coming from a new non-Hermitian parametrization of Scarf II potential could be the features of a general one-dimensional scattering potential ($V(\pm \infty)=0$). This opens up a scope for further investigations. It is also desired to investigate the possibility  of  more than two ({\it one}) spectral singularities in one-dimensional non-Hermitian ({\it complex PT-symmetric}) scattering potentials. 
 
\section*{References}
\begin{enumerate}
\item C. M. Bender and S. Boettcher 1998, Phys. Rev. Lett. {\bf 80}, 5243.
\item C. M. Bender 2007, Rep. Prog. Phys. {\bf 70} 947.
\item Z. H. Musslimani, K. G. Makris, R. El-Ganainy, and D.N. Christodoulides 2008,
Phys. Rev. Lett. {\bf 100}, 103904.\\ K. G. Makris, R. El-Ganainy, D.N. Christodoulides,
and Z. H. Musslimani 2008, Phys. Rev. Lett. {\bf 101}, 080402.\\
A. Guo, G. J. Salamo, D. Duchesne, R. Morandotti, M. Volatier-Ravat, V. Aimez, G. A. Siviloglou,
D. N. Christodoulides 2009, Phys. Rev. Lett. {\bf 103} 093902.
\item A. Mostafazadeh 2002, J. Math. Phys. {\bf 43} 205.
\item Z. Ahmed 2001, Phys. Rev. A 64, 042716 (2001).
\item R. N. Deb, A. Khare, B.D. Roy 2003, Phys. Lett A {\bf 307} 215.   
\item Z. Ahmed 2004, Phys. Lett. A {\bf 324} 152. 
\item A. Mostafazadeh and H. Mehr-Dehnavi 2009, J. Phys. A {\bf 42} 125303 (2009). \\
      A. Mostafazadeh 2009, Phys. Rev. Lett. {\bf 102} 220402.\\
      A. Mostafazadeh 2009, J. Phys. A: Math. Theor. {\bf 44} 375302.
\item B. F. Samsonov 2005, J. Phys. A: Math. Gen. {\bf 38} L571.    
\item H. F. Jones 1999, Phys. Lett. A {\bf 262} 242.\\
      J. K. Boyd 2001, J. Math. Phys. {\bf 42} 15.\\
      Z. Ahmed 2001, Phys,. Lett. A {\bf 286} 231.  
\item S. Longhi 2009, Phys. Rev. Lett. {\bf 103} 123609.
\item A. V. Sokolov, A.A. Andrianov, F. Cannata 2006, J.Phys. A: Math. Gen. 10207. 
\item Z. Ahmed 2001, Phys. Lett. A {\bf 282} 343; {\bf 287} 295.
\item Z. Ahmed 2009, J. Phys. A: Math. Theor. {\bf 42} 473005. 
\item B. Bagchi and C. Quesne 2000, Phys. Lett. A {\bf 273} 285.\\
      G. Levai and M. Znojil 2002,  J. Phys. A {\bf 35} 8795.\\
      C. S. Jia, S. C. Li, Y. Li and L. T. Sun 2002,  Phys. Lett. A {\bf 300} 115.\\
      G. Levai, F. Cannata and A. Ventura 2002,  Phys. Lett. A {\bf 300} 271.\\
      A. Sinha and R. Roychowdhury 2002,  Phys. Lett. A {\bf 301} 163.    
\item A. Khare and U. P. Sukhatme 1988, J. Phys. A: Math. Gen. {\bf 21} L501.
\item G. Levai, F. Cannata and A. Ventura 2001, J. Phys. A: Math. Gen. {\bf 34} 839.
\item B. Bagchi, C. Quesne 2010, J. Phys. A: Math. Theor. {\bf 43} 305301.
\end{enumerate} 

\begin{table}[h]

	\centering
		\begin{tabular}{|c|c|c|c|}
		\hline
		Hamiltonian & Reflectivity  & Transmitivity \\
		\hline
		$\{1\}$ & $R_{left}=R_{right}$, & $T_{left}=T_{right},$  \\
		Hermitian &  $R(-k)=R(k)$ &  $T(k)=T(-k)$ \\
		\hline
		$\{2\}$ Non-Hermitian  & $R_{left} = R_{right},$ & $T_{left}=T_{right},$  \\
		(P-symmetric) & $R(-k) \ne R(k)$ &  $T(-k) \ne T(k)$  \\
		\hline
		$\{3\}$ & $R_{left} \ne R_{right},$ & $T_{left}=T_{right},$ \\
		Non-Hermitian &  $R(-k) \ne R(k)$ &  $T(-k) \ne T(k)$ \\
		\hline
	  $\{4\}$ Non-Hermitian & $R_{left}\ne R_{right},$ & $T_{left}=T_{right},$ \\
		(PT-symmetric) &  $R_{left}(-k)=R_{right}(k)$ &  $T(-k)=T(k)$  \\
		\hline
				
		\end{tabular}

		\caption{Table displaying the proposed and existing (in)variances reflectivity and transmitivity.
		The proposal $\{3\}$ can be proved using the method of Ref.[5]. However, the result $\{4\}$ is a conjecture
		whose proof is desired.} 
		
\end{table}

\begin{figure}
\centering
\includegraphics[width=9 cm,height=15 cm]{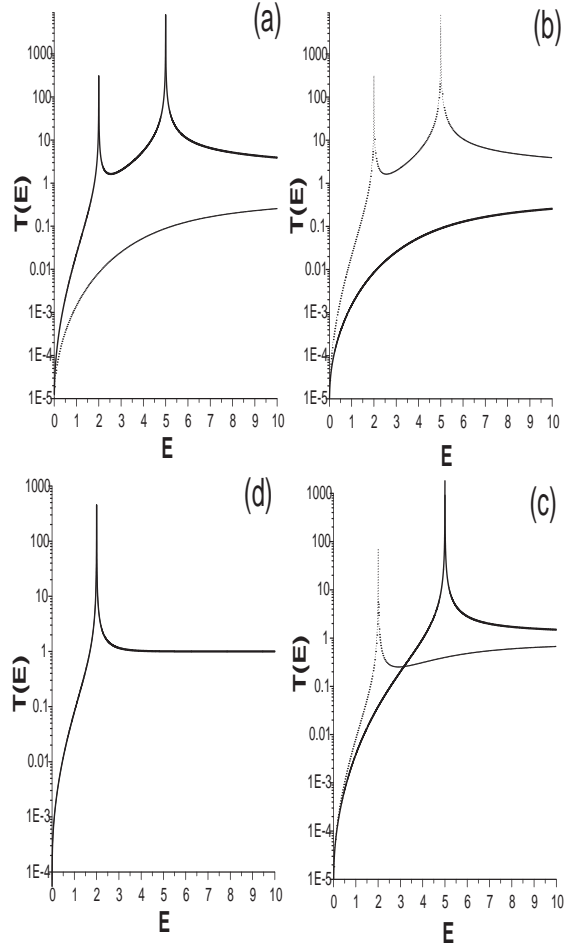}
\caption{ For complex scarf II potential [with the new parametrization (Eq.(10))], the transmitivity $T(E)$ is plotted as a function of energy $E$. 
The dark curve represents the $T(k)$ and the faint curve represents
the time-reversed transmitivity $T(-k)$. In (a) $\alpha=\sqrt{2}$ and $\beta= \sqrt{5}$, notice two peaks in the dark curve at $E=2,5$ and no 
peak in the faint one. In (b) $\alpha=-\sqrt{2}$ and $\beta= -\sqrt{5}$, notice two peaks in the faint  curve at $E=2,5$ and no peak in the dark curve.
In (c) $\alpha=-\sqrt{2}$ and $\beta= \sqrt{5}$, notice one peak in the faint  curve at $E=2$ and one peak in the dark curve at $E=5$. 
In (d) in the case when the potential becomes complex PT-symmetric $(\alpha=-\beta=\sqrt{2})$ both transmitivities coincide (the dark and the faint curves merge
together) and there is a single spectral singularity at $E=2$. Here we have taken $n=m=0$ (see Eq.(10)). The spectral singularities occur at $E=\alpha^2,\beta^2$.}
\end{figure}

\end{document}